\documentclass{Interspeech2024}

\interspeechcameraready 

\usepackage{multirow}
\usepackage{multicol}
\usepackage{makecell}
\usepackage{threeparttable}
\usepackage{tabularx}
\usepackage{xcolor}

\definecolor{xiaomi_gray}{HTML}{777777}

\newcommand{\cinnamon}{\textsc{Llama 2}\xspace}
\newcommand{\whisper}{\makecell{Whisper \\ (Large-v3)}}

\title{Enhancing Automated Audio Captioning via Large Language Models with Optimized Audio Encoding}

\name{Jizhong}{Liu}
\name[affiliation={\,\ast}]{Gang}{Li}
\name{Junbo}{Zhang}
\name[affiliation={\,\sharp}]{Heinrich}{Dinkel}
\name[affiliation={\,\sharp}]{Yongqing}{Wang}
\name[affiliation={\,\sharp}]{Zhiyong}{Yan}
\name{\newline Yujun}{Wang}
\name{Bin}{Wang}

\address{AI Lab, Xiaomi Corporation, China 
    \thanks{$^\ast$ Corresponding author.}
    \thanks{$^\sharp$ Equal contribution.}
    }
\email{\{liujizhong1,ligang5,zhangjunbo1\}@xiaomi.com}


\begin{document}

\maketitle

\begin{abstract}
Automated audio captioning (AAC) is an audio-to-text task to describe audio contents in natural language. Recently, the advancements in large language models (LLMs), with improvements in training approaches for audio encoders, have opened up possibilities for improving AAC. Thus, we explore enhancing AAC from three aspects: 1) a pre-trained audio encoder via consistent ensemble distillation (CED) is used to improve the effectivity of acoustic tokens, with a querying transformer (Q-Former) bridging the modality gap to LLM and compress acoustic tokens; 2) we investigate the advantages of using a \cinnamon with 7B parameters as the decoder; 3) another pre-trained LLM corrects text errors caused by insufficient training data and annotation ambiguities. Both the audio encoder and text decoder are optimized by low-rank adaptation (LoRA). Experiments show that each of these enhancements is effective. Our method obtains a 33.0 SPIDEr-FL score, outperforming the winner of DCASE 2023 Task 6A.
\end{abstract}

\footnotetext[1]{Available: \url{https://github.com/frankenliu/LOAE}}

\section{Introduction}

Automated audio captioning (AAC) is a multimodal task that aims to describe the audio content in natural language~\cite{review}. Distinct from speech-to-text conversion, AAC focuses on audio-to-text conversion to capture the underlying acoustic semantic information. AAC studies have gathered increasing interest in recent years, driven by the rising demand for intelligent interactions and information retrievals.

In recent studies, the typical encoder-decoder architecture has been progressively constructed~\cite{review}. The audio encoder extracts acoustic tokens from the input audio, while the text decoder generates the caption based on acoustic tokens. Generally, audio or speech extractors serve as the encoder (e.g., PANNs~\cite{panns}, ASTs~\cite{ast}, BEATs~\cite{beats}, SpeechT5~\cite{t5}, and Whisper~\cite{whisper}), and language models serve as the decoder (e.g., BERT~\cite{bert}, GPT-2~\cite{gpt2}, and BART~\cite{bart}). Despite various encoder-decoder combinations, state-of-the-arts (SOTAs) consistently leverage pre-trained models. For instance, the top-ranked method in the Detection and Classification of Acoustic Scenes and Events (DCASE) 2023 Challenge~\cite{top1} uses a BEATs-BART architecture. SALMONN~\cite{salmonn} focuses on interaction according to audio inputs, which utilize dual encoders and a large language model (LLM) decoder.

However, the existing AAC methods~\cite{review,top1,top2,top3,whisper_aac} suffer from the following common shortcomings: 1) The training of audio encoders may need to be sufficiently refined for AAC tasks. Too many generated acoustic tokens exacerbate decoding complexity. For example, captioning a 30-second audio clip by BEATs or Whisper yields approximately 1500 acoustic tokens. 2) Text decoders struggle to accurately describe long audio clips containing multiple overlapping events in natural language. A more powerful language model is required to understand and describe acoustic features effectively. 3) The training data is insufficient and imperfect, even using all available public datasets for AAC. The widely used datasets Clotho~\cite{clotho} and AudioCaps~\cite{audiocaps} are both crowdsourced, wherein many sound events are ambiguous, leading AAC decoders to learn incorrect descriptions. Current studies have failed to employ well-trained general-purpose models to address the shortage of training data.

To overcome these shortcomings, we propose a new method via \textbf{L}LMs with \textbf{O}ptimized \textbf{A}udio \textbf{E}ncoding (LOAE)\footnotemark[1]. The main contributions of this work are as follows:

\begin{itemize}
\item For audio encoding, we combine the consistent ensemble distillation (CED)~\cite{ced} encoder and low-rank adaptation (LoRA)~\cite{lora}, with better performance and much fewer output acoustic tokens. To bridge the modality gap, a fixed frame rate querying transformer (Q-Former) \cite{qformer} down-samples acoustic tokens of the audio encoder, simultaneously enhancing encoding attention and reducing decoding complexity.
\item For text decoding, a \cinnamon with 7B parameters \cite{llama} is fine-tuned with LoRA~\cite{lora} and an instruction prompt, which addresses the challenge of adapting LLMs to downstream tasks and boasts robust capabilities of text descriptions.
\item To overcome the insufficient training data and annotation ambiguities, another pre-trained LLM with the instruction prompt is used as the post-corrector, which provides additional language processing ability from the general-purpose model. For convenience, we directly use the ChatGPT-3.5~\cite{gpt3.5} application programming interfaces (APIs).
\end{itemize}

Experimental results demonstrate that each of these contributions is effective. Our method outperforms the winner of DCASE 2023 Task 6A~\cite{top1} on almost all metrics.

\section{Method}

\begin{figure}[t]
  \centering
  \includegraphics[width=2.2in]{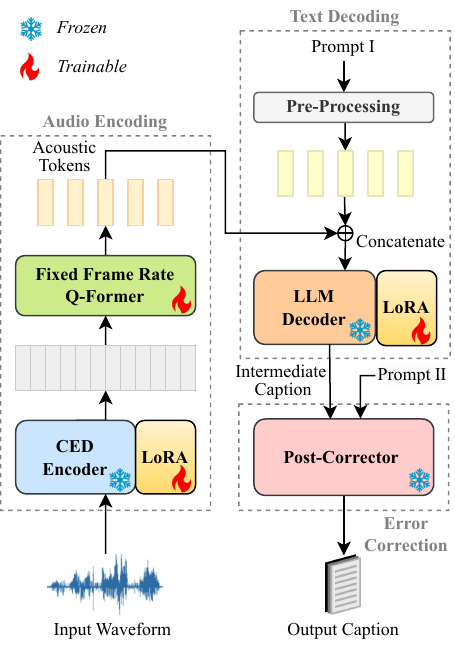}
  \caption{The architecture of the proposed LOAE method integrates an optimized encoding-decoding framework, with LoRA fine-tuning strategy. The LLM component remains frozen, while Q-Former bridges the encoder and the decoder. Additionally, an extra LLM serves as the post-corrector.}
  \label{fig:speech_production}
\end{figure}

The proposed method shares a similar overall architecture with current mainstream methods, maintaining a structure that combines audio encoding and text decoding, and the entire system is trained as a whole using cross-entropy loss~\cite{review}. Innovatively, we optimize the respective modules, as depicted in Figure \ref{fig:speech_production}. 

\textbf{Audio Encoding.} Our method leverages CED (base)~\cite{ced} as the audio encoder. CED ensembles multiple teacher models with a cumulative parameter count exceeding 1.2 billion and distilled to a student model with 85 million parameters. Unlike many audio encoders~\cite{ast, whisper} that pad inputs to fixed lengths, CED's improved positional embedding allows for better generalization across varying input lengths. CED significantly reduces computational overhead by utilizing 64-dimensional mel-filterbanks and employing $16 \times 16$ patches without overlap. A pre-trained CED model\footnotemark[2] without the output linear header serves as the audio encoder and is fine-tuned using LoRA \cite{lora}. To feed the encoded acoustic tokens to the LLM, every 17 tokens are compressed to 1 token using Q-Former~\cite{qformer}. The final number of acoustic tokens is equal to those processed by the widely used 14-layer convolutional neural network (CNN14) \cite{panns}. This down-sampling rate makes sure the fairness of the following ablation study.
\footnotetext[2]{\url{https://github.com/RicherMans/CED}}

\textbf{Text Decoding.} The \cinnamon~\cite{llama} with 7B parameters, which is also applied LoRA using the AAC training set, serves as the text decoder. This refined decoder can produce more precise captions by leveraging optimized audio encoding and the deeper understanding provided by LLMs. As described in Table \ref{tab:prompts} (\texttt{Prompt} \texttt{\uppercase\expandafter{\romannumeral 1}} in Figure \ref{fig:speech_production}), an instruction prompt guides \cinnamon in understanding AAC tasks.

\begin{table}[t!]
  \renewcommand{\arraystretch}{1}
  \footnotesize
  \caption{Instruction prompts of the proposed LOAE method.}
  \label{tab:prompts}
  \centering
  \begin{tabular}{cc}
    \toprule
    Prompt \uppercase\expandafter{\romannumeral 1} & \begin{minipage}{6cm} Describe the detail of this audio: $<$AcousticTokens$>$ $\backslash$n --- $\backslash$n Detailed: \end{minipage} \\
    \midrule
    Prompt \uppercase\expandafter{\romannumeral 2} & \begin{minipage}{6cm} Revise the sentence to make it more correct and idiomatic: $\backslash$n rain is falling on a tin roof ==$>$ rain is falling on the tin roof $\backslash$n $<$Text$>$ ==$>$ \end{minipage} \\
    \bottomrule
  \end{tabular}
\end{table}

\textbf{Error Correction.} Under current conditions of insufficient training data and annotation ambiguities, the AAC model may learn incorrect patterns, such as single phrase loops and grammatical errors. While data augmentation can mitigate errors to some extent, linguistic errors persist. Thus, a pre-trained LLM is employed as a post-corrector. Before applying the corrector, a grammatical Error Detector \cite{fense} estimates the error probability of the output caption text, and the corrector only activates when the error probability exceeds 0.90. To guide the error correction process, the instruction prompt (\texttt{Prompt} \texttt{\uppercase\expandafter{\romannumeral 2}}) contains a sentence that does not conform to linguistic conventions, and \texttt{<Text>} denotes the input sentence (intermediate caption in Figure \ref{fig:speech_production}) of the post-corrector.

\begin{table}[t!]
  \renewcommand{\arraystretch}{1}
  \scriptsize
  \begin{threeparttable}
  \caption{Performance comparisons on Clotho. The scores of SOTA methods are obtained from the official report of DCASE 2023 Task 6A\protect\footnotemark[6], where SPIDEr-FL (SD-F) is the ranking metric. Our scores are calculated from the evaluation set. Scores in {\color{xiaomi_gray}{gray}} are obtained with the help of ensemble strategies.}
  \label{tab:compare}
  \centering
  \begin{tabular}{cccccccc}
    \toprule
    ~                                                            & ME            & CD     & SP    & SD    & \textbf{SD-F} & SB            & FS            \\
    \midrule
    Wu \textit{et al.}~\cite{top1}                               & 19.5                      & 50.5          & \textbf{14.9} & 32.7          & 32.7          & 53.6          & 53.6          \\
    \color{xiaomi_gray}{Cho \textit{et al.}~\cite{top2}}         & \color{xiaomi_gray}{19.7} & \color{xiaomi_gray}{\textbf{53.9}} & \color{xiaomi_gray}{\textbf{14.9}} & \color{xiaomi_gray}{\textbf{34.4}} & \color{xiaomi_gray}{31.5} & \color{xiaomi_gray}{53.0} & \color{xiaomi_gray}{48.8} \\
    \color{xiaomi_gray}{Labb\'{e} \textit{et al.}~\cite{top3}}   & \color{xiaomi_gray}{19.3} & \color{xiaomi_gray}{48.6} & \color{xiaomi_gray}{14.2} & \color{xiaomi_gray}{31.4} & \color{xiaomi_gray}{31.4} & \color{xiaomi_gray}{52.3} & \color{xiaomi_gray}{52.2} \\
    Kadl{\v{c}}{\'\i}k \textit{et al.}~\cite{whisper_aac}        & 17.2          & 41.4  & 12.3   & 26.9  & 26.7          & 49.5          & 49.2          \\
    LOAE (Ours)                                                  & \textbf{19.7} & 51.3  & 14.7   & 33.0  & \textbf{33.0} & \textbf{53.8} & \textbf{53.8} \\
    \bottomrule
  \end{tabular}
  \end{threeparttable}
\end{table}

\begin{table*}[t!]
  \renewcommand{\arraystretch}{1}
  \scriptsize
  \caption{Comparison of audio encoding. The decoding strategy is \cinnamon with LoRA, and the post-corrector is not used.}
  \label{tab:encoder}
  \centering
  \begin{tabular}{ccccccccc|ccccccc}
    \toprule
                              &             & \multicolumn{7}{c}{Clotho}                                & \multicolumn{7}{c}{AudioCaps}                           \\
    \cmidrule(lr){3-16}
    Encoder                   & Strategy    & ME   & CD   & SP   & SD   & \textbf{SD-F} & SB   & FS    & ME   & CD   & SP   & SD   & \textbf{SD-F} & SB   & FS   \\
    \midrule
    \multirow{2}*{CNN14}      & Frozen      & 17.8 & 42.8 & 12.7 & 27.8 & 27.6          & 49.3 & 49.1  & 25.0 & 73.5 & 18.3 & 45.9 & 45.9          & 63.2 & 63.2 \\
                              & Fine-Tuning & 19.4 & 49.3 & 14.2 & 31.8 & 31.6          & 52.4 & 52.1  & 25.5 & 77.2 & 18.3 & 47.8 & 47.5          & 64.7 & 64.0 \\
    \midrule
    \multirow{3}*{BEATs}      & Frozen      & 18.8 & 47.0 & 13.6 & 30.3 & 30.2          & 52.0 & 51.7  & 25.6 & 78.0 & 19.0 & 48.5 & 48.4          & 66.0 & 65.7 \\
                              & Fine-Tuning & 19.3 & 49.0 & 14.3 & 31.7 & 31.3          & 53.1 & 52.7  & 25.9 & 76.7 & 18.9 & 47.8 & 47.7          & 65.9 & 65.6 \\
                              & LoRA        & 19.5 & 48.9 & 14.1 & 31.5 & 31.3          & 52.9 & 52.6  & 25.5 & 78.6 & 18.5 & 48.6 & 48.4          & 65.6 & 65.3 \\
    \midrule
    \multirow{3}*{\whisper}   & Frozen      & 18.7 & 47.2 & 13.6 & 30.4 & 30.1          & 51.4 & 50.9  & 24.7 & 71.8 & 17.7 & 44.8 & 44.8          & 63.0 & 62.9 \\
                              & Fine-Tuning & 18.3 & 45.3 & 13.3 & 29.3 & 29.1          & 51.3 & 51.0  & 24.0 & 71.1 & 17.1 & 44.1 & 44.0          & 62.3 & 62.2 \\
                              & LoRA        & 18.8 & 46.9 & 13.5 & 30.2 & 30.0          & 52.2 & 51.9  & 24.7 & 71.8 & 17.7 & 44.8 & 44.8          & 63.0 & 62.9 \\
    \midrule
    \multirow{3}*{\makecell{CED \\ (Base)}}        & Frozen      & 19.2 & 49.8 & 14.4 & 32.1 & 31.9          & 53.1 & 52.7  & 26.1 & 80.3 & 19.0 & 49.7 & 49.6          & 66.0 & 65.7 \\
                              & Fine-Tuning & 19.3 & 49.7 & 14.3 & 32.0 & 31.7          & 53.6 & 53.1  & 26.3 & 79.6 & 18.8 & 49.2 & 49.0          & 66.3 & 65.8  \\
                              & LoRA        & \textbf{19.7} & \textbf{51.4} & \textbf{14.7} & \textbf{33.1} & \textbf{32.8} & \textbf{53.8} & \textbf{53.2} & \textbf{26.7} & \textbf{81.6} & \textbf{19.3} & \textbf{50.5} & \textbf{50.4} & \textbf{66.4} & \textbf{66.2} \\
    \bottomrule
  \end{tabular}
\end{table*}

\begin{table*}[t!]
  \renewcommand{\arraystretch}{1}
  \scriptsize
  \caption{Comparison of text decoding. The encoding strategy is CED with LoRA and Q-Former.}
  \label{tab:decoder}
  \centering
  \begin{tabular}{ccccccccc|ccccccc}
    \toprule
                             &             & \multicolumn{7}{c}{Clotho}                                & \multicolumn{7}{c}{AudioCaps}                           \\
    \cmidrule(lr){3-16}
    Decoder                  & Strategy    & ME   & CD   & SP   & SD   & \textbf{SD-F} & SB   & FS    & ME            & CD   & SP            & SD   & \textbf{SD-F} & SB   & FS   \\
    \midrule
    \multirow{3}*{\makecell{BART \\ (Base)}}      & Frozen      & 15.9 & 36.6 & 11.5 & 24.0 & 23.8          & 50.3 & 49.8  & 18.9          & 54.8 & 14.0          & 34.4 & 33.2 & 55.9   & 53.8 \\
                             & Fine-Tuning & 18.2 & 45.6 & 13.0 & 29.3 & 29.0          & 52.2 & 51.4  & 25.0          & 77.3 & 18.3          & 47.8 & 47.5 & 65.1   & 64.5 \\
                             & LoRA        & 17.7 & 44.9 & 13.0 & 29.0 & 28.6          & 52.6 & 51.8  & 24.0          & 73.2 & 17.6          & 45.4 & 45.4 & 65.1   & 65.1 \\
    \midrule
    \multirow{3}*{\makecell{GPT-2 \\ (Large)}}     & Frozen      & 18.6 & 44.4 & 13.2 & 28.8 & 17.4          & 52.8 & 32.1  & 27.0          & 56.7 & 19.2          & 38.0 & 17.1 & 66.2   & 30.4 \\
                             & Fine-Tuning & 18.1 & 46.4 & 12.9 & 29.7 & 29.5          & 52.1 & 51.8  & 26.2          & 57.2 & 18.2          & 37.7 & 23.0 & 65.5   & 39.4 \\
                             & LoRA        & 18.4 & 47.0 & 13.3 & 30.1 & 30.1          & 52.8 & 52.5  & \textbf{27.2} & 60.4 & \textbf{19.6} & 40.0 & 21.3 & 66.0   & 35.5 \\
    \midrule
    \multirow{2}*{{\makecell{\cinnamon \\ (7B parameters)}}} & Frozen      & 19.0 & 47.8 & 13.7 & 30.7 & 30.4          & 52.9 & 52.2  & 25.5          & 79.5 & 18.8          & 49.1 & 49.0 & 65.6   & 65.4 \\
                             & LoRA        & \textbf{19.7} & \textbf{51.4} & \textbf{14.7} & \textbf{33.1} & \textbf{32.8} & \textbf{53.8} & \textbf{53.2} & 26.7 & \textbf{81.6} & 19.3 & \textbf{50.5} & \textbf{50.4} & \textbf{66.4} & \textbf{66.2} \\
    \bottomrule
  \end{tabular}
\end{table*}

\begin{table}[t!]
  \renewcommand{\arraystretch}{1}
  \scriptsize
  \caption{Ablation study for the post-corrector. The encoding strategy is CED with LoRA and Q-Former, and the decoding strategy is \cinnamon with LoRA.}
  \label{tab:postgpt}
  \centering
  \begin{tabular}{cccc}
    \toprule
                              &               & With Post-Corrector & W/O \\
    \midrule          
    \multirow{7}*{Clotho}     & ME            & 19.7                & 19.7 \\
                              & CD            & 51.3                & \textbf{51.4} \\
                              & SP            & 14.7                & 14.7 \\
                              & SD            & 33.0                & \textbf{33.1} \\
                              & \textbf{SD-F} & \textbf{33.0}       & 32.8 \\
                              & SB            & 53.8                & 53.8 \\
                              & FS            & \textbf{53.8}       & 53.2 \\
    \midrule
    \multirow{7}*{AudioCaps}  & ME            & 26.7                & 26.7 \\
                              & CD            & 81.6                & 81.6 \\
                              & SP            & 19.3                & 19.3 \\
                              & SD            & 50.5                & 50.5 \\
                              & \textbf{SD-F} & \textbf{50.5}       & 50.4 \\
                              & SB            & 66.4                & 66.4 \\
                              & FS            & \textbf{66.4}       & 66.3 \\
    \bottomrule
  \end{tabular}
\end{table}

\section{Experimental Setup}
\label{sec:setup}

\subsection{Datasets}
\label{sec:datasets}

Referring to the DCASE participating teams (e.g.,~\cite{top2, top3}), the AAC models are trained on three datasets: WavCaps~\cite{wavcaps}, Clotho~\cite{clotho}, and AudioCaps~\cite{audiocaps}. WavCaps is a large-scale weakly labeled audio captioning dataset comprising approximately 400k audio clips. These clips are paired with captions sourced from various sources, including AudioSet~\cite{audioset}, BBC sound effects\footnotemark[3], FreeSound\footnotemark[4], and SoundBible\footnotemark[5]. The captions in WavCaps are generated through a three-stage processing pipeline using ChatGPT~\cite{gpt3.5}. Clotho contains 4981 audio clips lasting 15 to 30 seconds, each annotated with five reference captions. AudioCaps comprises 46k audio clips lasting 10 seconds. In the AudioCaps training subset, each audio clip is annotated with one caption, while in the evaluation and test subsets, each clip is annotated with five captions.

The primary evaluation data is the Clotho dataset, which serves in the benchmark for ranking in the DCASE 2023 Challenge. Additionally, the AudioCaps dataset is included in ablation studies to enhance credibility. 

\footnotetext[3]{\url{https://sound-effects.bbcrewind.co.uk/}}
\footnotetext[4]{\url{https://freesound.org/}}
\footnotetext[5]{\url{https://soundbible.com/}}
\footnotetext[6]{\url{https://dcase.community/challenge2023/task-automated-audio-captioning-results}}

\subsection{Metrics}
\label{sec:metrics}
The experiments refer to all metrics presented in DCASE 2023 Task 6A, including METEOR~\cite{meteor}, CIDEr~\cite{cider}, SPICE~\cite{spice}, SPIDEr~\cite{spider}, SPIDEr-FL~\cite{fense}, Sentence-BERT~\cite{sbert}, and FENSE~\cite{fense}. The ranking metric of DCASE 2023 is SPIDEr-FL, which utilizes the Error Detector of FENSE to penalize the SPIDEr score of the sentence with an error probability greater than 90\%. METEOR and CIDEr are both based on n-gram overlap, while SPICE focuses on the overlap computed on semantic graphs constructed by objects, attributes, and relations. SPIDEr is the mean of CIDEr and SPICE. Sentence-BERT and FENSE are both BERT-based models to evaluate the similarity between the ground truth and the generated caption, with FENSE incorporating an Error Detector to penalize erroneous sentences. In Tables \ref{tab:compare} to \ref{tab:postgpt}, ME, CD, SP, SD, SD-F, SB, and FS denote METEOR, CIDEr, SPICE, SPIDEr, SPIDEr-FL, Sentence-BERT, and FENSE. All scores are scaled to a percentage scale by multiplying them by 100.

\subsection{Implementation Details}
\label{sec:imdetail}
The models are optimized on an AdamW optimizer~\cite{adamw} with a weight decay coefficient of $1\times10^{-6}$ and warming up first 2 epochs. We train the models using three datasets together with 15 epochs, a batch size of 48, and a learning rate of $5\times10^{-5}$, then we fine-tune the models separately on Clotho and AudioCaps with 30 epochs, a batch size of 32, and a learning rate of $5\times10^{-6}$. LoRA matrices are added to the ``q'' and ``v'' of Transformer~\cite{transformer}. The audio sampling rate is 16 kHz.

\begin{table}[t!]
  \renewcommand{\arraystretch}{1}
  \footnotesize
  \caption{Examples of error correction using post-corrector.}
  \label{tab:examples}
  \centering
  \begin{tabularx}{0.45\textwidth}{XX}
    \toprule
    \multicolumn{1}{c}{Input Text} & \multicolumn{1}{c}{Output Text}\\
    \midrule
    \begin{minipage}{\hsize} a car drives by and then another car drives by and then another car  drives by and then another car drives by and then another car drives by \end{minipage} &
    \begin{minipage}{\hsize} several cars drive past one after the other \end{minipage} \\
    \midrule
    \begin{minipage}{\hsize} a horse is galloping and a horse is neighing \end{minipage} &
    \begin{minipage}{\hsize} a horse gallops while neighing \end{minipage} \\
    \midrule
    \begin{minipage}{\hsize} a man speaks and a rooster crows \end{minipage} &
    \begin{minipage}{\hsize} a man is speaking and a rooster is crowing \end{minipage} \\
    \bottomrule
  \end{tabularx}
\end{table}

\section{Results}

\subsection{Overall Performance Comparison}
The proposed LOAE method is compared with the top three methods~\cite{top1, top2, top3} as well as  an interesting method~\cite{whisper_aac} from DCASE 2023 Task 6A. Specifically, Wu \textit{et al.}~\cite{top1} construct a BEATs-BART model with Conformer~\cite{conformer} and ChatGPT mix-up augmentation~\cite{top1}, Cho \textit{et al.}~\cite{top2} construct a CNN14-BART model with AL-MixGen augmentation~\cite{AL-MixGen}, Labb\'{e} \textit{et al.}~\cite{top3} construct a ConvNeXt-Transformer~\cite{ConvNeXt, transformer} model, and Kadl{\v{c}}{\'\i}k \textit{et al.}~\cite{whisper_aac} directly train an ASR model (Whisper large-v2) using AAC datasets.

The comparative results are given in Table \ref{tab:compare}, which presents all available SOTA scores. LOAE ranks first on the official metric SPIDEr-FL by a single model, and almost all scores of LOAE are higher than the DCASE 2023 winner (Wu \textit{et al.}). Even compared to the DCASE 2023 runner-up (Cho \textit{et al.}) who uses an ensemble strategy, our method still outperforms on learning-based metrics, i.e., LOAE is more in line with semantic similarities than simple mathematical similarities. The results of Kadl{\v{c}}{\'\i}k \textit{et al.} indicate that the automatic speech recognition (ASR) model cannot perform AAC tasks well. Hence, AAC remains a challenging independent study.

\subsection{Audio Encoding Comparison}
This experiment compares different audio encoding strategies, including the pre-trained CNN14, BEATs, Whisper, and CED. CNN14 is not equipped with Q-Former due to the limited number of output tokens, and the case of CNN14 with LoRA is excluded due to its non-Transformer structure. Table \ref{tab:encoder} presents detailed results on the Clotho and AudioCaps datasets, showcasing the impact of freezing, fine-tuning, and applying LoRA to the audio encoders. Although AudioCaps is a subset of AudioSet, CNN14, BEATs and CED are all pre-trained with AudioSet, ensuring relative fairness.

CED consistently outperforms other encoders in terms of SPIDEr-FL scores. The fine-tuned CED, benefiting from LoRA, demonstrates enhanced performance, achieving a remarkable SPIDEr-FL score of 32.8 on Clotho and 50.4 on AudioCaps.
Despite extracting 64-channel Mel spectrograms, which is half the quantity used by BEATs and Whisper, CED still obtains higher scores than comparative encoders. This result highlights the effectiveness of CED in reducing decoding complexity while maintaining high fidelity in generated captions. The ability of CED to handle varying input lengths without padding contributes to its superior generalization across diverse audio clips.
Although their studies~\cite{salmonn, whisperat} have reported Whisper excels as a general audio encoder, the comparison result show that the model pre-trained on speech datasets may not perform AAC tasks effectively.

\subsection{Text Decoding Comparison}

This experiment compares three different decoders, including BART, GPT-2, and \cinnamon. The case of trainable \cinnamon is excluded due to efficiency issues. As shown in Table \ref{tab:decoder}, \cinnamon achieves much higher scores, indicating that LLMs indeed enhance text decoding capability. LoRA provides significant assistance for LLMs adapting to downstream tasks, thereby avoiding the challenges of full fine-tuning. 

The superior performance of \cinnamon  can be attributed to its substantial model size and the nuanced optimization by LoRA fine-tuning strategy. The larger model capacity of \cinnamon allows for a more profound understanding of complex acoustic features, enabling it to generate more contextually relevant and accurate captions. Furthermore, the LoRA fine-tuning approach contributes to the decoder's adaptability to the specific nuances of the AAC task, refining its parameters to better align with the intricacies of the dataset. This combination of model size and targeted fine-tuning enables \cinnamon to outperform other decoders in decoding long audio clips with multiple overlapping events, showcasing its effectiveness in capturing and expressing the rich acoustic information in the input audio. Moreover, the SPIDEr-FL scores of GPT-2 are abnormally low.  By observing the generated captions, it is found that GPT-2 excessively prioritizes lexical matching, leading to frequent grammar errors in short sentences of AudioCaps.

\subsection{Impact of LoRA}

LoRA is a crucial optimization technique applied to both audio encoders and text decoders. Table \ref{tab:encoder} and Table \ref{tab:decoder} highlight the impact of LoRA on the performances of different components.

The audio encoding comparison of frozen, fine-tuning, and LoRA reveals that CED with LoRA outperforms other methods across various metrics. Compared to full fine-tuning, LoRA preserves the efficiency gains associated with training resources and yields superior performance. The remarkable effectiveness of LoRA lies in its capacity to seamlessly adapt the pre-trained audio encoder to the unique demands of AAC, while simultaneously retaining the advantages of the original pre-training. Similarly, in text decoding, LoRA consistently enhances the decoder's performance across multiple evaluation metrics compared to the frozen counterpart.
Our investigation proves that LoRA is a practical and effective optimization technique for both audio encoding and text decoding. Of course, LoRA may be less effective in some structures, such as BEATs.

\subsection{Post-Corrector Analysis}

Table \ref{tab:postgpt} presents an ablation study for the post-corrector component of LOAE. The results indicate that the post-corrector improves performance on both datasets. With the post-corrector activated, there is a certain enhancement in SPIDEr-FL scores, demonstrating its effectiveness in addressing linguistic errors caused by insufficient training data and annotation ambiguities.

Table \ref{tab:examples} provides three typical examples of the error correction. The post-corrector can rectify phrase loop errors, optimize the linguistic expression of parallel events, correct typical grammatical errors, etc. These errors may stem from the lack of training data and annotation ambiguities. Although LLM decoders have significantly fewer text errors than traditional decoders, the error correction mechanism is still beneficial under the current dataset conditions.

However, it should be noted that the post-corrector's impact varies across metrics, with SPIDEr-FL showcasing the most substantial improvement. This result suggests that the post-corrector is particularly effective in refining the overall fluency and correctness of generated captions, contributing to the impressive SPIDEr-FL scores achieved by LOAE.

\section{Conclusion}
\maketitle

In this work, we comprehensively explore enhancing AAC by addressing critical challenges in audio encoding and text decoding. An optimized audio encoding, specifically the CED encoder with LoRA, refines the accuracy of acoustic tokens. Our investigation of decoding capabilities focuses on implementing a LLM-based decoder, specifically using a \cinnamon model with 7B parameters. Q-Former connects the audio encoder and the LLM-based text decoder, building an effective bridge to capture and represent the underlying acoustic features while reducing decoding complexity. Furthermore, our results highlight LoRA's importance, demonstrating its resource-efficient training and enhanced performance in the AAC task. LoRA poses a valuable optimization strategy for AAC, which finely adapts pre-trained models to the nuances in audio encoding and text decoding. Lastly, the post-corrector based on a pre-trained LLM and an instruction prompt is proven beneficial for the inference stage. This post-corrector safeguards against errors from insufficient training data and annotation ambiguities, further improving the overall captioning accuracy. 
The proposed method outperforms the winner of DCASE 2023 Task 6A and obtains a 33.0 SPIDEr-FL score. The ablation studies indicate that all components are crucial to our architecture.

Future work will aim to improve the accuracy and efficiency of applying LLMs. For instance, we may explore combining the text decoding and error correction within a single LLM.

\newpage 
\bibliographystyle{IEEEtran}

\end{document}